\documentclass{elsart}

\usepackage{epsfig}
\usepackage{amssymb}

\begin{document}
\begin{frontmatter}

\title{``Madelung model'' prediction for dependence of 
lattice parameter on nanocrystal size.}

\author{Vasili Perebeinos\corauthref{cor}} 
\ead{vasili@bnl.gov}
\ead[url]{http://cmth.phy.bnl.gov/$\sim$perebein}
\corauth[cor]{Corresponding author.}
\address{Department of Physics, Brookhaven National Laboratory, Upton,
New York 1973-5000}

\author{Siu-Wai Chan} and
\author{Feng Zhang}
\address{Department of Applied Physics and Applied Mathematics,
Columbia University, New York, NY 10027}


\begin{abstract}
The competition between the long range Coulomb attractive 
and the short range repulsive interaction in ionic nanocrystals 
creates an effective negative pressure, which causes the
lattice parameter $a$ to increase with decreasing nanoparticle size $d$. 
A simple ``Madelung model'' is used to predict the dependence of the 
lattice parameter for CeO$_2$ and BaTiO$_3$ on $d$, $\delta a/a=\alpha/d$, 
for  $\alpha=$0.22 \AA \ and 0.18 \AA \ respectively. 
The model predictions are compared with experimental results. 
\end{abstract}

\end{frontmatter}

How properties of ionic crystals change
when their surface to volume ratio 
becomes large is an active area of 
research \cite{Zhang,Spanier,Tsunekawa2,Tsunekawa1,Tsunekawa3,Sayle,Cordatos}. 
The ceria nanoparticles CeO$_2$ have been studied experimentally using 
Raman, X-ray diffraction, photoabsorption, high resolution transmission 
electron microscopy (HRTEM) \cite{Zhang,Spanier},
and transmission electron microscopy \cite{Tsunekawa2,Tsunekawa1,Tsunekawa3}.
The fascinating property of ionic nanoparticles is the lattice parameter  
increase with nanoparticle decreasing size \cite{Zhang,Tsunekawa1,Tsunekawa3}. 
Intuitively one would expect the lattice parameter  to decrease as a result 
of surface stress as was observed in gold nanoparticles \cite{Mays}. 


Nanocrystal properties are significantly 
different from their bulk counterparts. Previous theoretical treatments 
\cite{Sayle,Cordatos} used realistic potentials for small Ce$_n$O$_{2n}$ 
cluster calculations. It was shown that flourite structure for ceria 
was not achieved for $n$ smaller than $50$. In the present work we investigate
lattice parameter dependence on a cluster size for $d=4-20$ nm or 
$n=10^3-10^5$. 
Charge neutral octahedra nanocrystals are surface terminated with oxygen 
anion layers along the eight $\{111\}$ faces. 
The balance between the long range Coulomb and the short range atomic 
interactions leads to the optimal lattice parameter  $a_0$ dependent on 
the size of the cluster. 
The volume energy per formula unit in the bulk material is usually modeled as
\begin{eqnarray}
\varepsilon_{\rm v}=\frac{BV_0}{2}\left(\frac{V}{V_0}-1\right)^2,
\label{Ev}
\end{eqnarray}
where $B$ is the bulk modulus and  $E_{\rm v}=N\varepsilon_{\rm v}$ is the 
volume energy of $N$ formula units of CeO$_2$, each filling volume space 
$V_0=a_0^3/4$. 
In the infinite crystal Eq. (\ref{Ev}) is the total free energy which 
is minimum at the bulk lattice parameter  $a_0$. 
In the nanocrystal the balance between the Coulomb and short range 
forces is altered due to the surface. 
The surface Coulomb energy  
creates a negative pressure which leads to the lattice parameter  expansion 
in the nanocrystal in accord with observation \cite{Zhang}. 

\begin{figure}
\psfig{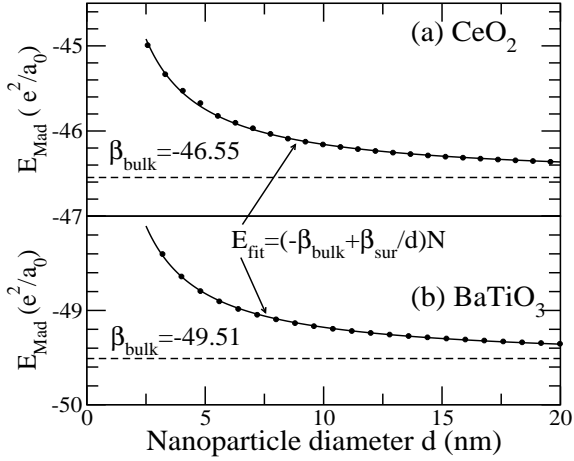}
\caption{Closed circles are model calculations of the Madelung energy 
on ionic (a) CeO$_2$ and (b) BaTiO$_3$ nanocrystals. 
The CeO$_2$ nanocrystal was terminated by 
eighth (111) planes and BaTiO$_3$ by six (100) planes. 
The solid lines are the best fits using Eq. (\protect{\ref{Emad}}).}
\label{fig1}
\end{figure}

To calculate the Coulomb energy in ionic nanoparticles we use the same 
procedure as Madelung energy calculation in bulk material \cite{Kittel}. 
In the fluorite (CaF$_2$) structure the Ce$^{4+}$ ions occupy 
a fcc sublattice and  O$^{2-}$ ions occupy the tetrahedral 
interstitial sites forming a simple cubic sublattice of length $a_0/2$. 
This gives 
a Madelung energy per formula unit 
$\varepsilon_{\rm Mad}=\beta_{\rm bulk}e^2/a$, 
where $\beta_{\rm bulk}=-46.55$.
The value of $\beta=Mz_iz_j/r_0$ 
is related to the Madelung constant $M=2.52$ for the 
CaF$_2$ crystal structure \cite{Lide}, where ionic charges are
$z_i=2$, $z_j=4$ and the nearest neighbor distance is $r_0=\sqrt{3}/4$ 
in units of $a$.
To calculate the surface Madelung energy we use a finite size cluster of 
Ce$^{4+}$ and O$^{2-}$ ions terminated by the oxygen layers along the 
eight $(111)$ planes. 
The surface energy is proportional to the area of  
nanoparticle $\sim N^{2/3}$. 
The diameter $d=(3N/2\pi)^{1/3}a_0$ is 
related to the number of Ce atoms $N$ such that the sphere of equal diameter 
contains an equal volume $NV_0$. 
The Madelung energy 
of a cluster (with a uniform lattice parameter) 
can be calculated for different diameters. The results are
shown on Fig. \ref{fig1}(a), which are well fitted by
\begin{eqnarray}
\frac{{\rm E}_{\rm Mad}}{N}=\left(-\beta_{\rm bulk}+
\frac{\beta_{\rm sur}}{d}\right)\frac{e^2}{a},
\label{Emad}
\end{eqnarray}
with two parameters $\beta_{\rm bulk}=46.55$ and $\beta_{\rm sur}=4.13$ nm.

%

The  uniform lattice parameter  expansion reduces the surface Madelung energy
creating the stress in the nanocrystal estimated with Eq. (\ref{Ev}). 
The total energy per Ce atom on the nanoparticle is
\begin{eqnarray}
\frac{{\rm E}_{\rm tot}}{N}=\frac{BV_0}{2}\left(\frac{a^3}{a_0^3}-1\right)^2
+\frac{\beta_{\rm sur}}{d}\frac{e^2}{a}.
\label{Emad2}
\end{eqnarray}
Here the first term is the volume contribution, which includes the 
Madelung energy of the infinite crystal, and the second term is the 
surface Coulomb energy from Eq. \ref{Emad}. 
Minimization of the total energy with respect 
to the lattice parameter  for fixed $N$ gives
\begin{eqnarray}
\frac{\delta a}{a_0}=\frac{1}{9BV_0}\frac{e^2}{a_0}\frac{\beta_{\rm sur}}{d}
=\frac{\alpha}{d}.
\label{res}
\end{eqnarray}

\begin{figure}
\psfig{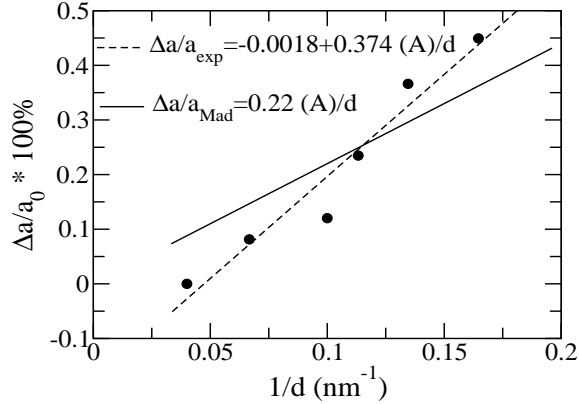}
\caption{Filled circles are experimental lattice parameters versus inverse 
diameter of CeO$_2$ nanoparticle.
Dashed line is the best fit for the experimental data 
\protect\cite{Zhang}. Solid line is a Madelung model prediction 
$\delta a/a=0.22$ \AA $/d$.}
\label{fig2}
\end{figure}

The bulk modulus $B=230\pm10$ GPA has been measured by high-pressure 
x-ray diffraction \cite{Duclos}. 
Taking $a_0=5.409$ \AA \ we obtain the relation $\delta a/a_0=0.22$ \AA$/d$
for the lattice parameter expansion shown on 
Fig. \ref{fig2}. The experimental lattice parameter  
(also shown on Fig. \ref{fig2}) expands with a larger slope $\alpha=0.37$ \AA.
Our model predicts 60\% of the measured value of $\alpha$. 
The missing 40\% of the lattice parameter  expansion 
might be assigned to the formation of point defects.
These defects are oxygen vacancies and simultaneous reduction of two 
ions of Ce$^{4+}$ to Ce$^{3+}$ for each oxygen vacancy formed,
maintaining charge neutrality. 
Electron spectroscopy \cite{Mullins} and 
X-ray photoelectron spectroscopy \cite{Tsunekawa2} show the presence of a
Ce$^{3+}$ component in nanocrystals.
Since Ce$^{3+}$ has a larger radius compared to Ce$^{4+}$ there may
be a lattice expansion accompanying the 
replacement of (Ce$^{4+}$)$_2$O$^{2-}$ by Ce$^{3+}V$ where $V$ is 
a neutral oxygen vacancy.
An upper limit in the Ce$^{3+}$ contribution to expansion 
was estimated \cite{Zhang} by interpolation of published 
results of the variation in ceria lattice parameter  
with RE$^{3+}$ radii from doping with different rare earth oxides 
(RE$_2$O$_3$) \cite{rev,McBride}. 
Recent neutron diffraction measurements on nano-scale powder of CeO$_2$ 
\cite{Egami} suggest associated oxygen defect formation 
(involving oxygen vacancies and self-interstitials) in
the fluorite structure. 
Such defects are not included in our model and may alter 
the surface Madelung energy estimation. 

\vspace{1.0pc}

The Madelung model is valid for any ionic nanocrystal.
We apply it for the BaTiO$_3$ nanoparticles to calculate
the competition between the long range Coulomb and the short range 
atomic interactions. The results are shown on Fig.\ref{fig1}(b). 
Ionic charges Ba$^{2+}$, Ti$^{4+}$ and O$^{2-}$ were used to calculate 
the surface energy of the nanoparticle. 
To preserve charge neutrality the cubic nanocrystal with size $d$ 
was terminated by the three BaO and three TiO$_2$ $\{100\}$ planes. 
The calculated value of $\beta_{\rm sur}=3.6$ nm along with
the experimental lattice parameter  $a_0=3.996$ \AA \ and bulk modulus
$B=196$ GPa \cite{Fischer} predicts $\delta a/a_0=0.18$ \AA $/d$. 
Existing data \cite{Tsunekawa3} on BaTiO$_3$ nanoparticles 
are not sufficient to verify the relationship 
Eq. (\ref{res}) and further experiments are needed.

In conclusion we propose a mechanism for lattice 
parameter  expansion of ionic nanocrystals 
due to the effective negative Madelung pressure. 
Using the experimental bulk modulus we quantitatively explain a
lattice parameter  expansion in ionic materials. 
A discrepancy between the model prediction and observed effect in CeO$_2$ 
can be attributed to the Ce$^{3+}$ component \cite{Tsunekawa2,Tsunekawa1}
and other defects not included in our model. 
The Ce$^{3+}$ concentration deduced from magnetic measurements 
\cite{Tsunekawa1} suggests a much smaller value than expected from 
lattice expansion measurements in accord with our theory.
A non-uniform lattice parameter  expansion 
and structural defect formation \cite{Egami} would 
change the quantitative answer, but not the conclusion that 
an effective negative Madelung pressure is significant in ionic 
nanocrystals.

{\bf Acknowledgments}

We are grateful to Philip B. Allen for critical reading of the manuscript, 
Michael Weinert for valuable discussions, and Stephen O'Brien for 
taking our attention on BaTiO$_3$ nanocrystals.
This work was supported in part by DOE  Grant No.\ DE-AC-02-98CH10886.
S.-W.C. and F.Z. acknowledge support by the MRSEC program of the National 
Science Foundation under DMR-9809687.

\end{document}